
\magnification=1200\overfullrule=0pt\baselineskip=14.5pt
\vsize=23truecm \hsize=15.5truecm \overfullrule=0pt\pageno=0

\font\titlefont=cmbx10 scaled \magstep1
\font\sectnfont=cmbx8  scaled \magstep1
%
%

\def\mname{\ifcase\month\or January \or February \or March \or April
           \or May \or June \or July \or August \or September
           \or October \or November \or December \fi}
\def\date{\hbox{\strut\mname \number\year}}
\def\banner{\hfill\hbox{\vbox{\offinterlineskip\crnum}}\relax}
\def\manner{\hbox{\vbox{\offinterlineskip\crnum\date}}
               \hfill\relax}
\footline={\ifnum\pageno=0\manner\else\hfil\number\pageno\hfil\fi}
%
%
%
\newcount\FIGURENUMBER\FIGURENUMBER=0
\def\FIG#1{\expandafter\ifx\csname FG#1\endcsname\relax
               \global\advance\FIGURENUMBER by 1
               \expandafter\xdef\csname FG#1\endcsname
                              {\the\FIGURENUMBER}\fi}
\def\figtag#1{\expandafter\ifx\csname FG#1\endcsname\relax
               \global\advance\FIGURENUMBER by 1
               \expandafter\xdef\csname FG#1\endcsname
                              {\the\FIGURENUMBER}\fi
              \csname FG#1\endcsname\relax}
\def\fig#1{\expandafter\ifx\csname FG#1\endcsname\relax
               \global\advance\FIGURENUMBER by 1
               \expandafter\xdef\csname FG#1\endcsname
                      {\the\FIGURENUMBER}\fi
           Fig.~\csname FG#1\endcsname\relax}
\def\figand#1#2{\expandafter\ifx\csname FG#1\endcsname\relax
               \global\advance\FIGURENUMBER by 1
               \expandafter\xdef\csname FG#1\endcsname
                      {\the\FIGURENUMBER}\fi
           \expandafter\ifx\csname FG#2\endcsname\relax
               \global\advance\FIGURENUMBER by 1
               \expandafter\xdef\csname FG#2\endcsname
                      {\the\FIGURENUMBER}\fi
           figures \csname FG#1\endcsname\ and
                   \csname FG#2\endcsname\relax}
\def\figto#1#2{\expandafter\ifx\csname FG#1\endcsname\relax
               \global\advance\FIGURENUMBER by 1
               \expandafter\xdef\csname FG#1\endcsname
                      {\the\FIGURENUMBER}\fi
           \expandafter\ifx\csname FG#2\endcsname\relax
               \global\advance\FIGURENUMBER by 1
               \expandafter\xdef\csname FG#2\endcsname
                      {\the\FIGURENUMBER}\fi
           figures \csname FG#1\endcsname--\csname FG#2\endcsname\relax}
\newcount\TABLENUMBER\TABLENUMBER=0
\def\TABLE#1{\expandafter\ifx\csname TB#1\endcsname\relax
               \global\advance\TABLENUMBER by 1
               \expandafter\xdef\csname TB#1\endcsname
                          {\the\TABLENUMBER}\fi}
\def\tabletag#1{\expandafter\ifx\csname TB#1\endcsname\relax
               \global\advance\TABLENUMBER by 1
               \expandafter\xdef\csname TB#1\endcsname
                          {\the\TABLENUMBER}\fi
             \csname TB#1\endcsname\relax}
\def\table#1{\expandafter\ifx\csname TB#1\endcsname\relax
               \global\advance\TABLENUMBER by 1
               \expandafter\xdef\csname TB#1\endcsname{\the\TABLENUMBER}\fi
             Table \csname TB#1\endcsname\relax}
\def\tableand#1#2{\expandafter\ifx\csname TB#1\endcsname\relax
               \global\advance\TABLENUMBER by 1
               \expandafter\xdef\csname TB#1\endcsname{\the\TABLENUMBER}\fi
             \expandafter\ifx\csname TB#2\endcsname\relax
               \global\advance\TABLENUMBER by 1
               \expandafter\xdef\csname TB#2\endcsname{\the\TABLENUMBER}\fi
             Tables \csname TB#1\endcsname{} and
                    \csname TB#2\endcsname\relax}
\def\tableto#1#2{\expandafter\ifx\csname TB#1\endcsname\relax
               \global\advance\TABLENUMBER by 1
               \expandafter\xdef\csname TB#1\endcsname{\the\TABLENUMBER}\fi
             \expandafter\ifx\csname TB#2\endcsname\relax
               \global\advance\TABLENUMBER by 1
               \expandafter\xdef\csname TB#2\endcsname{\the\TABLENUMBER}\fi
            Tables \csname TB#1\endcsname--\csname TB#2\endcsname\relax}
\newcount\REFERENCENUMBER\REFERENCENUMBER=0
\def\REF#1{\expandafter\ifx\csname RF#1\endcsname\relax
               \global\advance\REFERENCENUMBER by 1
               \expandafter\xdef\csname RF#1\endcsname
                         {\the\REFERENCENUMBER}\fi}
\def\reftag#1{\expandafter\ifx\csname RF#1\endcsname\relax
               \global\advance\REFERENCENUMBER by 1
               \expandafter\xdef\csname RF#1\endcsname
                      {\the\REFERENCENUMBER}\fi
             \csname RF#1\endcsname\relax}
\def\ref#1{\expandafter\ifx\csname RF#1\endcsname\relax
               \global\advance\REFERENCENUMBER by 1
               \expandafter\xdef\csname RF#1\endcsname
                      {\the\REFERENCENUMBER}\fi
             [\csname RF#1\endcsname]\relax}
\def\refto#1#2{\expandafter\ifx\csname RF#1\endcsname\relax
               \global\advance\REFERENCENUMBER by 1
               \expandafter\xdef\csname RF#1\endcsname
                      {\the\REFERENCENUMBER}\fi
           \expandafter\ifx\csname RF#2\endcsname\relax
               \global\advance\REFERENCENUMBER by 1
               \expandafter\xdef\csname RF#2\endcsname
                      {\the\REFERENCENUMBER}\fi
             [\csname RF#1\endcsname--\csname RF#2\endcsname]\relax}
\def\refand#1#2{\expandafter\ifx\csname RF#1\endcsname\relax
               \global\advance\REFERENCENUMBER by 1
               \expandafter\xdef\csname RF#1\endcsname
                      {\the\REFERENCENUMBER}\fi
           \expandafter\ifx\csname RF#2\endcsname\relax
               \global\advance\REFERENCENUMBER by 1
               \expandafter\xdef\csname RF#2\endcsname
                      {\the\REFERENCENUMBER}\fi
            [\csname RF#1\endcsname,\csname RF#2\endcsname]\relax}
\def\refs#1#2{\expandafter\ifx\csname RF#1\endcsname\relax
               \global\advance\REFERENCENUMBER by 1
               \expandafter\xdef\csname RF#1\endcsname
                      {\the\REFERENCENUMBER}\fi
           \expandafter\ifx\csname RF#2\endcsname\relax
               \global\advance\REFERENCENUMBER by 1
               \expandafter\xdef\csname RF#2\endcsname
                      {\the\REFERENCENUMBER}\fi
            [\csname RF#1\endcsname,\csname RF#2\endcsname]\relax}
\def\refss#1#2#3{\expandafter\ifx\csname RF#1\endcsname\relax
               \global\advance\REFERENCENUMBER by 1
               \expandafter\xdef\csname RF#1\endcsname
                      {\the\REFERENCENUMBER}\fi
           \expandafter\ifx\csname RF#2\endcsname\relax
               \global\advance\REFERENCENUMBER by 1
               \expandafter\xdef\csname RF#2\endcsname
                      {\the\REFERENCENUMBER}\fi
           \expandafter\ifx\csname RF#3\endcsname\relax
               \global\advance\REFERENCENUMBER by 1
               \expandafter\xdef\csname RF#3\endcsname
                      {\the\REFERENCENUMBER}\fi
            [\csname RF#1\endcsname,\csname
                RF#2\endcsname,\csname RF#3\endcsname]\relax}
\def\Ref#1{\expandafter\ifx\csname RF#1\endcsname\relax
               \global\advance\REFERENCENUMBER by 1
               \expandafter\xdef\csname RF#1\endcsname
                      {\the\REFERENCENUMBER}\fi
             Ref.~\csname RF#1\endcsname\relax}
\def\Refs#1#2{\expandafter\ifx\csname RF#1\endcsname\relax
               \global\advance\REFERENCENUMBER by 1
               \expandafter\xdef\csname RF#1\endcsname
                      {\the\REFERENCENUMBER}\fi
           \expandafter\ifx\csname RF#2\endcsname\relax
               \global\advance\REFERENCENUMBER by 1
               \expandafter\xdef\csname RF#2\endcsname
                      {\the\REFERENCENUMBER}\fi
        Refs.~\csname RF#1\endcsname{},\csname RF#2\endcsname\relax}
\def\Refto#1#2{\expandafter\ifx\csname RF#1\endcsname\relax
               \global\advance\REFERENCENUMBER by 1
               \expandafter\xdef\csname RF#1\endcsname
                      {\the\REFERENCENUMBER}\fi
           \expandafter\ifx\csname RF#2\endcsname\relax
               \global\advance\REFERENCENUMBER by 1
               \expandafter\xdef\csname RF#2\endcsname
                      {\the\REFERENCENUMBER}\fi
            Refs.~\csname RF#1\endcsname--\csname RF#2\endcsname]\relax}
\def\Refand#1#2{\expandafter\ifx\csname RF#1\endcsname\relax
               \global\advance\REFERENCENUMBER by 1
               \expandafter\xdef\csname RF#1\endcsname
                      {\the\REFERENCENUMBER}\fi
           \expandafter\ifx\csname RF#2\endcsname\relax
               \global\advance\REFERENCENUMBER by 1
               \expandafter\xdef\csname RF#2\endcsname
                      {\the\REFERENCENUMBER}\fi
        Refs.~\csname RF#1\endcsname{} and \csname RF#2\endcsname\relax}
\newcount\EQUATIONNUMBER\EQUATIONNUMBER=0
\def\EQ#1{\expandafter\ifx\csname EQ#1\endcsname\relax
               \global\advance\EQUATIONNUMBER by 1
               \expandafter\xdef\csname EQ#1\endcsname
                          {\the\EQUATIONNUMBER}\fi}
\def\eqtag#1{\expandafter\ifx\csname EQ#1\endcsname\relax
               \global\advance\EQUATIONNUMBER by 1
               \expandafter\xdef\csname EQ#1\endcsname
                      {\the\EQUATIONNUMBER}\fi
            \csname EQ#1\endcsname\relax}
\def\EQNO#1{\expandafter\ifx\csname EQ#1\endcsname\relax
               \global\advance\EQUATIONNUMBER by 1
               \expandafter\xdef\csname EQ#1\endcsname
                      {\the\EQUATIONNUMBER}\fi
            \eqno(\csname EQ#1\endcsname)\relax}
\def\EQNM#1{\expandafter\ifx\csname EQ#1\endcsname\relax
               \global\advance\EQUATIONNUMBER by 1
               \expandafter\xdef\csname EQ#1\endcsname
                      {\the\EQUATIONNUMBER}\fi
            (\csname EQ#1\endcsname)\relax}
\def\eq#1{\expandafter\ifx\csname EQ#1\endcsname\relax
               \global\advance\EQUATIONNUMBER by 1
               \expandafter\xdef\csname EQ#1\endcsname
                      {\the\EQUATIONNUMBER}\fi
          Eq.~(\csname EQ#1\endcsname)\relax}
\def\eqand#1#2{\expandafter\ifx\csname EQ#1\endcsname\relax
               \global\advance\EQUATIONNUMBER by 1
               \expandafter\xdef\csname EQ#1\endcsname
                        {\the\EQUATIONNUMBER}\fi
          \expandafter\ifx\csname EQ#2\endcsname\relax
               \global\advance\EQUATIONNUMBER by 1
               \expandafter\xdef\csname EQ#2\endcsname
                      {\the\EQUATIONNUMBER}\fi
         Eqs.~(\csname EQ#1\endcsname) and (\csname EQ#2\endcsname)\relax}
\def\eqto#1#2{\expandafter\ifx\csname EQ#1\endcsname\relax
               \global\advance\EQUATIONNUMBER by 1
               \expandafter\xdef\csname EQ#1\endcsname
                      {\the\EQUATIONNUMBER}\fi
          \expandafter\ifx\csname EQ#2\endcsname\relax
               \global\advance\EQUATIONNUMBER by 1
               \expandafter\xdef\csname EQ#2\endcsname
                      {\the\EQUATIONNUMBER}\fi
          Eqs.~\csname EQ#1\endcsname--\csname EQ#2\endcsname\relax}
\def\APEQNO#1{\expandafter\ifx\csname EQ#1\endcsname\relax
               \global\advance\EQUATIONNUMBER by 1
               \expandafter\xdef\csname EQ#1\endcsname
                      {\the\EQUATIONNUMBER}\fi
            \eqno(\APPENDIXNUMBER.\csname EQ#1\endcsname)\relax}
\def\APEQNM#1{\expandafter\ifx\csname EQ#1\endcsname\relax
               \global\advance\EQUATIONNUMBER by 1
               \expandafter\xdef\csname EQ#1\endcsname
                      {\the\EQUATIONNUMBER}\fi
            (\APPENDIXNUMBER.\csname EQ#1\endcsname)\relax}
\def\apeq#1{\expandafter\ifx\csname EQ#1\endcsname\relax
               \global\advance\EQUATIONNUMBER by 1
               \expandafter\xdef\csname EQ#1\endcsname
                      {\the\EQUATIONNUMBER}\fi
          Eq.~(\APPENDIXNUMBER.\csname EQ#1\endcsname)\relax}
%
\newcount\SECTIONNUMBER\SECTIONNUMBER=0
\newcount\SUBSECTIONNUMBER\SUBSECTIONNUMBER=0
\def\appendix#1#2{\global\let\APPENDIXNUMBER=#1\bigskip\goodbreak
     \line{{\sectnfont Appendix \APPENDIXNUMBER.\ #2}\hfil}\smallskip}
\def\section#1{\global\advance\SECTIONNUMBER by 1\SUBSECTIONNUMBER=0
      \bigskip\goodbreak\line{{\sectnfont \the\SECTIONNUMBER.\ #1}\hfil}
      \smallskip}
\def\subsection#1{\global\advance\SUBSECTIONNUMBER by 1
      \bigskip\goodbreak\line{{\sectnfont
         \the\SECTIONNUMBER.\the\SUBSECTIONNUMBER.\ #1}\hfil}
      \smallskip}
%

\def\NP{{\sl Nucl.\ Phys.\ }}
\def\PL{{\sl Phys.\ Lett.\ }}
\def\PR{{\sl Phys.\ Rev.\ }}

\def\PRL{{\sl Phys.\ Rev.\ Lett.\ }}

\vsize=23.5truecm\hsize=16.0truecm
\def\tifrnum{\hbox{TIFR/TH/94-29 \strut}}
\def\banner{\hfill\hbox{\offinterlineskip\vbox{\tifrnum}}\relax}
\def\manner{\hbox{\vbox{\offinterlineskip\bigskip\bigskip
                        \tifrnum\date}}\hfill\relax}
%
%

\def\N{{\scriptscriptstyle N}}
\def\S{{\scriptscriptstyle S}}
\def\T{{\scriptscriptstyle T}}
\newbox\slashbox\setbox\slashbox=\hbox{/}
\newdimen\slashdimen\slashdimen=\wd\slashbox
\def\lb{\hfil\penalty-10000}

\def\etc{{\sl etc.\/}}
\def\ie{{\sl i.e.\/}}

\def\ALPHAS{\alpha_\S}\def\alphas{\ifmmode\ALPHAS\else$\ALPHAS$\fi}
\def\SQRTS{\sqrt S}\def\sqrts{\ifmmode\SQRTS\else$\SQRTS$\fi}
\def\PT{p_\T}\def\pt{\ifmmode\PT\else$\PT$\fi}
\def\QT{q_\T}\def\qt{\ifmmode\QT\else$\QT$\fi}
\def\QTB{{\bf q}_\T}\def\Qt{\ifmmode\QTB\else$\QTB$\fi}
\def\PTS{/\kern-\slashdimen p_\T}\def\pts{\ifmmode\PTS\else$\PTS$\fi}
\def\QTS{/\kern-\slashdimen q_\T}\def\qts{\ifmmode\QTS\else$\QTS$\fi}
\def\PTBS{/\kern-\slashdimen {\bf p}_\T}\def\Pts{\ifmmode\PTBS\else$\PTBS$\fi}
\def\QTBS{/\kern-\slashdimen {\bf q}_\T}\def\Qts{\ifmmode\QTBS\else$\QTBS$\fi}
\def\PTCUT{{p_\T^{\rm cut}}}\def\ptcut{\ifmmode\PTCUT\else$\PTCUT$\fi}
\def\DIST{{\cal P}_\N(\Pts)}\def\dist{\ifmmode\DIST\else$\DIST$\fi}
\def\MULT{dN_\pi/dy}\def\mult{\ifmmode\MULT\else$\MULT$\fi}
\def\TI{T_0}\def\ti{\ifmmode\TI\else$\TI$\fi}
\def\GeV{${\rm GeV}^2$}
{\vsize=21truecm\banner\bigskip\baselineskip=15pt
\begingroup\titlefont\obeylines
\hfil JET MISSING TRANSVERSE MOMENTA\hfil
\hfil FOR QUARK-GLUON PLASMA THERMOMETRY\hfil
\endgroup\bigskip\bigskip
\bigskip\centerline{Sourendu Gupta}
\centerline{Theory Group, TIFR, Homi Bhabha Road, Bombay 400005, India.}
\bigskip\bigskip
\centerline{\bf ABSTRACT}\medskip
At the LHC, where hard multi-jet events are common, large imbalances in
the total jet-\pt{} can arise due to the interaction of jets with the
plasma formed in heavy-ion collisions. We find that this missing-\pt{}
spectrum yields a measurement of the initial temperature of the plasma.
Using commonly accepted guesses for this temperature, and ISAJET
estimates for jet cross sections, we predict an order of thousand events
at LHC with unbalanced $\pt>50$ GeV.
\vfil\eject}

For the heavy-ion collider mode of the LHC, with $\sqrts=5.5$ TeV, the
cross section for multi-jet events in which none of the jets is less
energetic than, say, 50 GeV is of the order of several $\mu{\rm b}$. In a
year's run one expects to see about $10^6$--$10^7$ jet events. Of these
about $10^4$ events contain four or more jets. The expected distribution
in the number of jets, $N$, is shown in \fig{isajet}. Since the underlying
event comes with a multiplicity density, \mult, of 2000--3000 pions per
unit of rapidity, the transverse energy within a cone of $\Delta R=0.5$
is about 20--30 GeV (assuming $\langle\pt\rangle\approx0.5$ GeV per pion).
Thus 50 GeV jets are easily observed above this background.

\vskip8truept\hrule\midinsert\vskip5.5truecm
\includegraphics{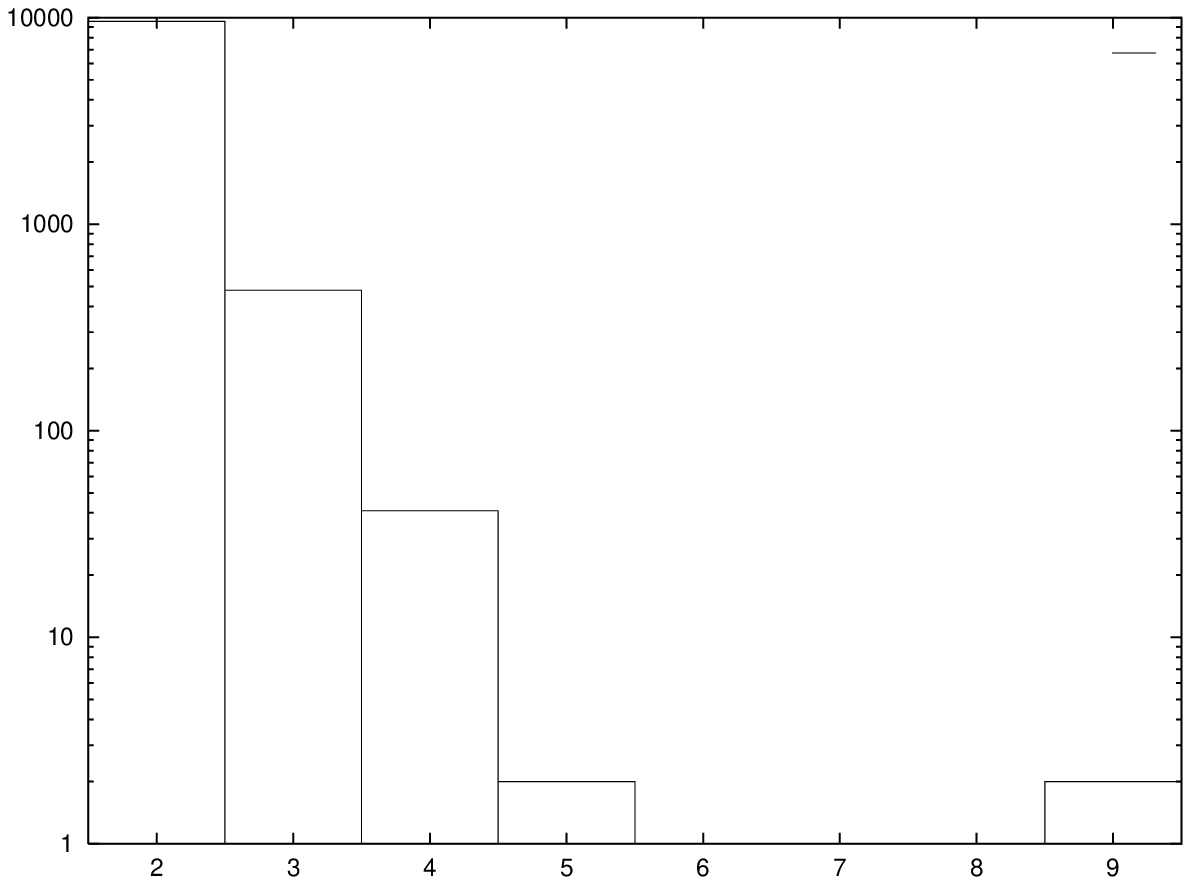}
\endinsert\centerline{Figure \figtag{isajet}}
The distribution in the number of jets obtained from a sample of $10^5$
events generated in ISAJET with the kinematics mentioned in the text.
\vskip8truept\hrule\vskip8truept

If simultaneously, a quark-gluon plasma is formed, then the evolution of
the jet is coupled to the plasma through strong interactions. This opens
up the possibility of measuring plasma properties using jets. Such
measurements could truly be called hard probes of the plasma. The simplest
measurements are kinematic, involving changes in the 4-momentum of jets as
they evolve through the plasma. It should be emphasised that these are
strictly {\it probes} and not {\it signals} of the plasma.

There is a history of related studies for dijet events. The first suggestion
\ref{appel} was to study acoplanarity in 2-jet events. A static model of the
plasma was used to show that small acoplanarities can exist. This was extended
in \ref{larry} to include the effects of hydrodynamics and a mixed phase. In
\ref{heinz} it was shown that a hadron gas could give numerically similiar
results. Acoplanarity is an useful variable only for 2-jet events. Very
recently energy-energy correlations in 2-jets events have also been
investigated \ref{gale}. A parallel
line of attack is to study the energy loss of a test parton evolving through
the plasma. This gives rise to the observation of jet quenching \ref{miklos}.

In the context of forthcoming LHC experiments, we believe that the simplest
measurement is of missing transverse momentum, \pts. This is a standard tool
for many other kinds of physics, and one of the LHC detectors can easily
perform this measurement. Such a measurement need not be restricted to dijet
events. If an experiment with $4\pi$ coverage observes exactly $N$-jets, with
a given \pts, then this measurement can be converted into a statement about
the initial temperature, \ti, of the plasma. The experimental analysis is
made robust by three pleasant aspects of the signal.
\item{\romannumeral1)}
   The \pts-spectrum is almost independent of the initial jet energies when
   each jet is harder than 50 GeV. Thus there is no need to guess the
   energy of the jet at the hard vertex, before its interaction with the
   plasma.
\item{\romannumeral2)}
   The \pts-spectra for different $N$ are related. Taken together, they
   provide a cross check for the physical origin of the missing momentum.
   Unlike acoplanarity, this measurement is not restricted to the dijet
   sample.
\item{\romannumeral3)}
   Using well-accepted guesses for \ti{} at LHC, we estimate about a
   thousand events with $\pts>50$ GeV. This would yield an accurate
   experimental measurement of \ti.

\noindent
Backgrounds coming from lack of full $4\pi$ coverage, contamination by the
underlying event, and the production of neutrinos and non-standard-model
particles can be made small, and will be discussed at the end of this letter.

In this letter we outline the computation of the \pts-spectrum,
and show the relation between
the \pts-spectra for different number of jets. After this we proceed
to display our results. We find that the \pts-spectrum is insensitive
to the initial energy in the jet, but strongly sensitive to \ti.
It is to be emphasised that \pts{} arises when considering only the
jets. If the \pt{} of all hadrons in each collision were to be added
up, there would be no \pts.

{\bf The Computation:}\
The partons which give rise to jets are produced through a hard
scattering in the earliest times of heavy-ion collisions. In the absence
of a medium, the measurement of the momenta of the particles constituting
the jet would give the momenta of these partons. This enables us to compute
jet production cross sections in perturbative QCD. In the presence of a
medium, final state scattering would occur. If the medium is a quark
gluon plasma, then the individual scatterings may again be computed in
perturbative QCD.

The energy of the constituents of the plasma are typically of the order of
a GeV, whereas we shall
consider jets carrying energy of order 50 GeV. The difference in the
scales translates to the kinematic statement that the
fractional momentum change of the jet in each scattering is small. Hence,
the individual scatterings may be considered independent. We can then write
the \Pts-spectrum in an $N$-jet process as the convolution of many
re-scattering probabilities, $F_i$,---
$$\dist\;\equiv\;{1\over\sigma}{d^2\sigma\over d^2\pts}
  \;=\;\sum_{n_1,\cdots,n_N} \prod_{\alpha=1}^N {1\over n_\alpha!}
    \int\left\{\prod_\mu d^2{\bf q}^\mu_\alpha F({\bf q}_\mu^\alpha)\right\}
     \delta^2\left(\Pts-\sum_{\mu,\alpha}{\bf q}_\mu^\alpha\right).
                 \EQNO{dist}$$
Here the cross section $\sigma$ may be considered to be differential in
as many other variables as required. The individual distributions, $F_i$,
may also be considered to be functions of these other variables. Note that
the distribution in \pts{} is solely given in terms of the interaction
between the jet and the plasma. The details of the hard scattering process,
including the perturbative expansion of the matrix elements, possible
resummations of initial state radiation, jet definitions, \etc, are all
subsumed into the factors of $\sigma$.

The probability of generating a missing transverse momentum \Qts{} for a hard
parton of species $j$ is given by
$$F^{(j)}(\Qt)\;=\;\sum_i\int_{\tau_0}^{\tau_f}d\tau {\cal N}_i(\tau)
      f_i^{(j)}(\Qts),         \EQNO{prob} $$
where the index $i$ runs over the species of partons in the plasma and
${\cal N}_i(\tau)$ is the number density of partons of species $i$ at time
$\tau$. The hard parton is assumed to interact with the plasma for times
$\tau_0\le\tau\le\tau_f$. The initial time, $\tau_0$, will be taken as the
thermalisation time of the plasma. The final time, $\tau_f$, can be
obtained from kinematic considerations. The function $f_i^{(j)}$ contains
the singular part of the invariant cross section for the
scattering of partons of species $i$ and $j$. Since the relevant jets are
gluonic, in the rest of this paper we shall lighten the notation by dropping
the index $j$.

A two-dimensional Fourier transform decouples the
$\delta$-function in \eq{dist} and yields the characteristic
function of the \pts-distribution as the exponential of the characteristic
functions of the distribution $F$ \ref{parisi}.
Thus, after normalisation, we obtain
$$S_\N(b)\;=\;\int d^2\Pts \exp\bigl[-i{\bf b}\cdot\Pts\bigr]
       \dist\;=\;\exp[F_\N(b)], \EQNO{defcf}$$
where
$$F_\N(b)\;=\;2\pi\sum_{\alpha=1}^N
   \sum_i\int_{\tau_0}^{\tau_f^\alpha}d\tau{\cal N}_i(\tau)
       \int_\mu^{\sqrt{2E^\alpha T(\tau)}} dq q f_i(q)
             \left[J_0(bq)-1\right]. \EQNO{deffb}$$
We have used the fact that $f_i$ is a function only of the magnitude of
\Qts{} to perform the angular integration in the Fourier transform and obtain
the Bessel function $J_0(b\qts)$. The subtraction in \eq{deffb} makes
$F(0)=0$, \ie, $S(0)=1$, and hence takes care of the normalisation of \dist.
The lower limit, $\mu$, of the integral over $q$ regulates a leftover
logarithmic divergence, and is chosen to be the Debye screening mass
evaluated at one-loop order.
The \pts-distribution is recovered by an inverse Fourier transform, which,
after the angular integration is performed, takes the form
$$\dist\;=\;{1\over2\pi}\int db b J_0(b\pts) S_\N(b). \EQNO{distexp}$$
Note that many variables are implicit in this formula. The dynamics of
the plasma enters through the distributions ${\cal N}_i$ and explicitly
also through $\tau_0$. The kinematics of the jets enter through the
quantities $E^\alpha$ and $\tau_f^\alpha$. We specify these dependences
next.

We begin with the evolution of the plasma. In this letter we assume the
plasma to be described by longitudinal boost-invariant hydrodynamics. We
shall take as our initial conditions a head-on collision of two nuclei of
mass number $A$. The plasma volume is then cylindrically symmetric, with
radius $R=r_0A^{1/3}$ ($r_0=6\ {\rm GeV}^{-1}$). A local temperature may
be defined, $T(\tau)$. The particle distributions, as functions of this
temperature, have an expansion in \alphas. To leading order, they are
$${\cal N}_i(\tau)\;=\; g_i T^3(\tau),\qquad{\rm where}\qquad
	g_i\;=\; \cases{{16\zeta(3)\over\pi^2},&for $i=g$\cr
	        {4N_f\zeta(3)\over\pi^2},&for $i=q,\bar q$\cr}
	\EQNO{fft} $$
The evolution of the local temperature keeps the entropy density fixed.
Hence
$$\tau T^3(\tau)\;=\;\tau_0\ti^3.  \EQNO{bjevol}$$
\ti{} may be related to the pion multiplicity density assuming
that there is no entropy generation during hadronisation of the plasma.
We can write
$$\tau_0\ti^3\;=\;{\pi\over3\zeta(3)R^2}{dN_\pi\over dy}. \EQNO{t0}$$
In most of our calculations, we shall use this formula to specify \ti{}
in terms of the multiplicity. However, in actual experiments, the jet
measurement would, among other things, check the validity of this formula.

The functions $f_i$ also have expansions in $\alphas$. Again we retain
the leading term, and obtain
$$f_i(\qt)\;=\;h_i{\alphas^2(\qt)\over\qt^4},\qquad{\rm where}\qquad
	h_i\;=\;\cases{{9\over2},& for $i=g$\cr
	    2,& for $i=q,\bar q$\cr} \EQNO{cross}$$
{}From \eqand{fft}{cross} we obtain
$$F(\qt,\tau)\;=\;\lambda{\alphas^2(\qt)\over\qt^4},
           \qquad{\rm where}\qquad
    \lambda={36(4+N_f)\zeta(3)\over\pi^2}T^3(\tau).   \EQNO{static}$$
We have used the one-loop order expression for $\alphas(\qt)$.

Next, we define the kinematics of the multijet event. Assume the hard
vertex to be at distance $r$ from the axis of symmetry. The momentum of
the $\alpha$-th parton can be taken to be
$$p^\alpha\;=\; E^\alpha(1,{\rm sech}y^\alpha\cos\theta^\alpha,
       {\rm sech}y^\alpha\sin\theta^\alpha,\tanh y^\alpha), \EQNO{mom}$$
A simple computation now yields
$$\tau_f^\alpha\;=\;\sqrt{r^2+R^2-2rR\cos\phi},\qquad{\rm where}\qquad
   \sin(\phi-\theta^\alpha)=-{r\over R}\sin\theta^\alpha.   \EQNO{kinem}$$

Since the polar angles and the position of the hard vertex, $r$, are not
measured, these variables must be averaged over. This is the final
ingredient in the computation.
We assume that the cross section for a hard process is just
the convolution of the hard cross section with a nuclear density function.
Assuming an uniform density, we obtain
$${\cal P}_\N(\pts)\;=\;{1\over2\pi}\int db b J_0\bigl(\pts b\bigr)
     \int{drr\sqrt{R^2-r^2}d\Omega\over\Omega R^3} S_\N(b).  \EQNO{final}$$
The volume element $(d\Omega/\Omega)$ averages over the jet angles
$\theta^\alpha$ ($\alpha=1,\cdots,N$) with uniform measure. The
distribution is clearly a function of the energies and rapidities of the
jets. The \pts-spectrum is now completely specified.

{\bf Some features:}\
A few remarks may help understand the results of the numerical work
reported later. In this approach to the interaction between jets
and a plasma, each jet undergoes successive scatterings from the
soft plasma independently of the others. The net effect is a large
imbalance in \pt. This remaining \pt{} is found in the soft hadronic
background underlying the jets, but is invisible in the jet measurement.

Although each single collision of the jet with the
plasma only changes the vectorial \pt{} by a number of the order of
$T$, the number of collisions along the path is large. It is
proportional to ${\cal N}$ integrated along the path, and hence
increases at $\ti^3$.
Note that $\langle\pts^2\rangle$ computed from ${\cal P}_1(\pts)$
equals that from $F(\pts)$. Resummation only changes the shape of
the \pts{} distribution, introducing a longer tail at the cost of
the small-\pts{} region.

In \ref{appel} QCD initial state radiation was also resummed {\sl a la}
Collins and Soper \ref{cs} and added to the \pts-distribution. We have
not done this for a simple reason. Such a resummation is performed for
an $N$-jet inclusive cross section. For large \pts{} this estimates the
effect of events where the number of hard jets is larger than $N$. Since
we consider exclusive $N$-jet events, and large \pts, inclusion of such
a term in our formul\ae{} would be erroneous.

In \ref{larry} a component of the scattering from the mixed phase was
also included. It was observed in \ref{gale} that jets escape the plasma
before the mixed phase can form. The kinematics in \eq{kinem} also lead to
the same conclusion. Hence we have not included mixed phase cross sections
in our computations. A further ingredient in some studies of the plasma
is the inclusion of transverse hydrodynamics. While this decreases the
plasma lifetime, it also enhances the plasma volume. Thus the net effect
on the \pts-spectrum may be small. It will be studied in more detail in
a later paper.

{\bf Results:}\
For orientation we display results for a single jet travelling
through a plasma. This may arise, for example, as the balancing
jet for high-\pt{} Drell-Yan events. In this case \pts{} is the
imbalance in \pt{} between the jet and the lepton pair. For $N=1$
$d\Omega/\Omega=d\theta/2\pi$. We take $A=200$, $\tau_0=0.1$ and
the jet rapidity to be zero. For $\mult=2000$, the \pts-spectrum
is shown in \fig{onejet}. We obtain $\langle\pts^2\rangle=16$ \GeV,
and find that the distribution goes to zero rapidly near $\pts=40$
GeV. The spectra for jet energies in the range 100 to 1000 GeV are
almost equal. The dependence on $A$ at fixed \ti{} is small;
$\langle\pts^2\rangle$ changes by 6\% in going from $A=50$ to $A=200$.
Since this change is due only to the thickness of the plasma firetube,
it also demonstrates that we can neglect changes in \pts-distributions
for small changes in the impact parameter of the colliding nuclei.
Increasing \mult{} to 3000 broadens the spectrum a little;
$\langle\pts^2\rangle=18$ \GeV{} in this case. However, it is
unlikely that such small \pts{} can be reliably measured above the
soft hadronic background.

\vskip8truept\hrule\midinsert\vskip5.5truecm
\includegraphics{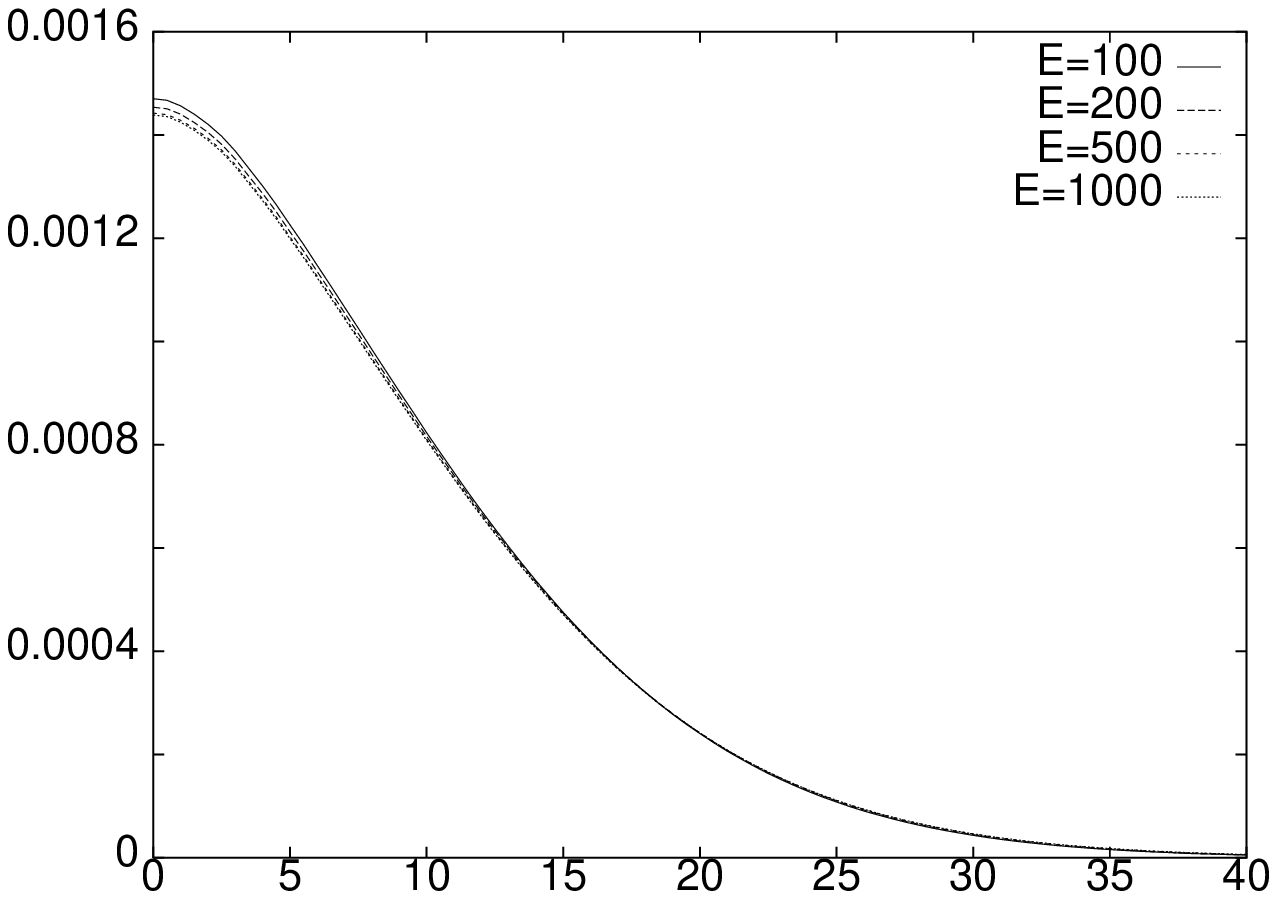}
\endinsert\centerline{Figure \figtag{onejet}}
\centerline{${\cal P}_1(\pts)$ plotted against \pts{} (in GeV)
 for $\mult=2000$.}

\vskip8truept\hrule\midinsert\vskip5.5truecm
\includegraphics{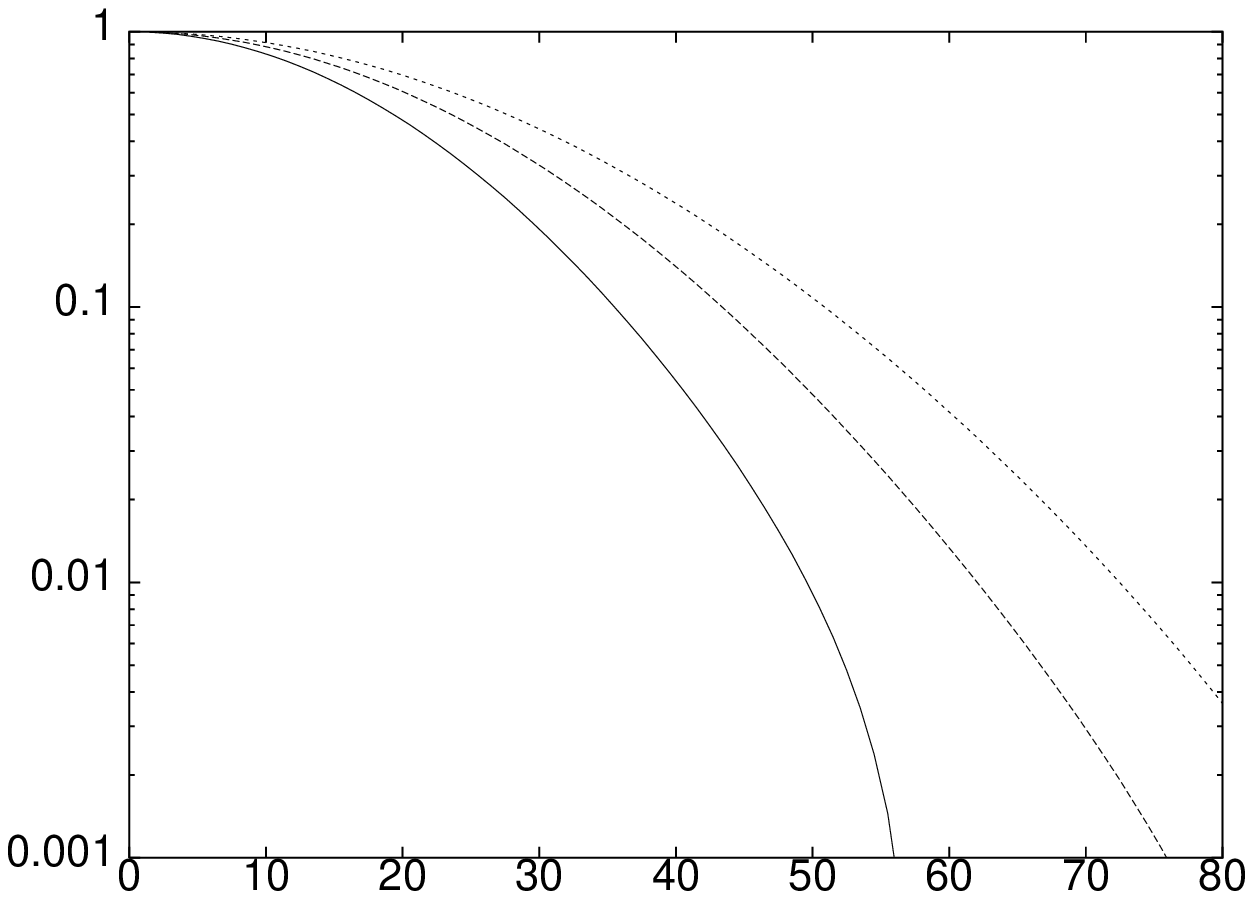}
\includegraphics{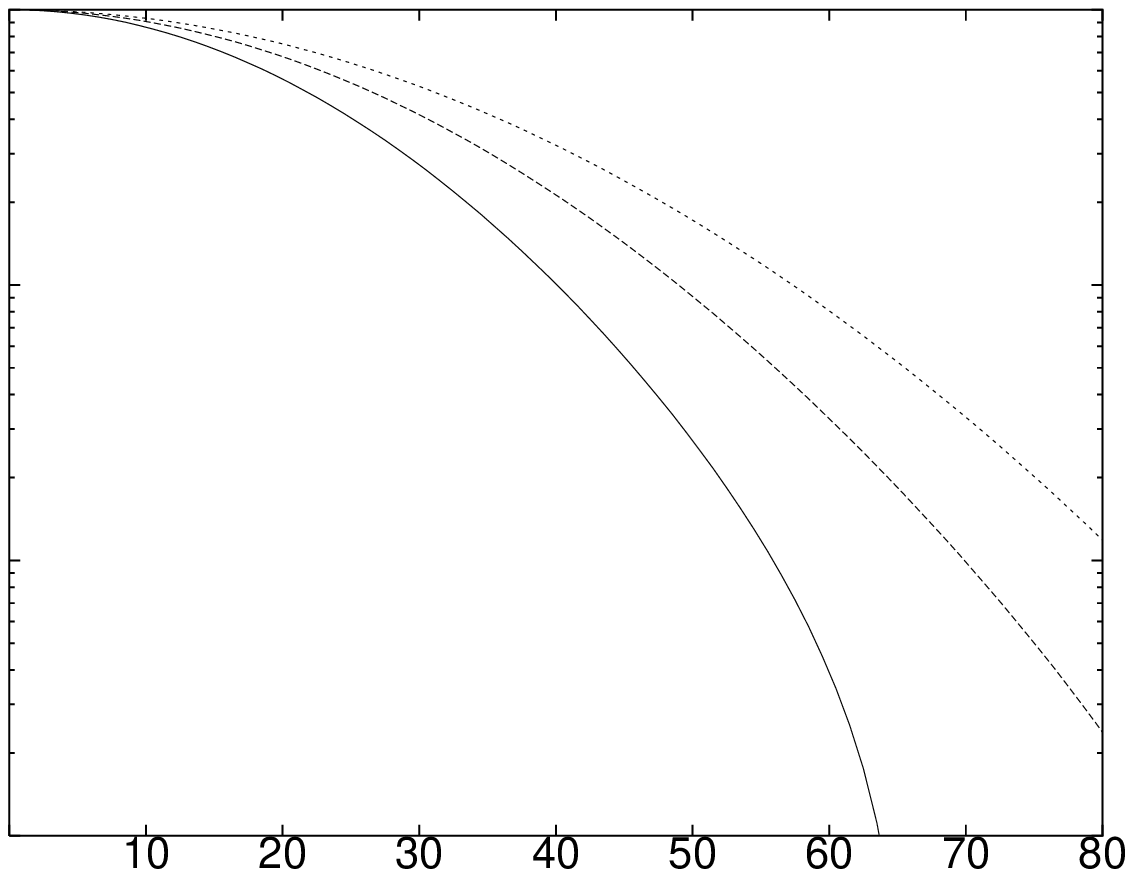}
\endinsert\centerline{Figure \figtag{ncrease}}
${\cal P}^>_\N(\ptcut)$ plotted against \ptcut{} (in GeV) for multiplicity
densities of (a) 2000 and (b) 3000, per unit of rapidity. The curves are for
2 (full line), 3 (dashed line) and 4 (dotted line) jets at zero rapidity
with $\sqrt{\hat s}=400$ GeV.
\vskip8truept\hrule\vskip8truept

Multijet events display a much harder \pts-spectrum. At fixed $\mult=
2000$, the values of $\langle\pts^2\rangle$ are 23 \GeV, 29 \GeV{} and
33 \GeV{} for $N=2$, 3 and 4, respectively. In \fig{ncrease} we display
$${\cal P}^>_\N(\ptcut)\;=\;1-\int_0^\ptcut d\pts^2\dist, \EQNO{frac}$$
\ie, the fraction of events with $\pts>\ptcut$. For $\mult=2000$, we
see that the fraction of events surviving a cut of $\pts>50$ GeV are
1\%, 5\% and 10\% of the total. For comparison, the corresponding numbers
increase to 3\%, 10\% and 17\% respectively, when $\mult=3000$.
Thus, the distinction between different values of \ti{} is easily
observable, and becomes easier with increasing $N$.

{\bf Backgrounds:}\
The main background is the uncertainty in \pts{} measurements due to the
high multiplicity in the underlying event. If the transverse energy within
a jet cone of $\Delta R=0.5$ due to this effect is $E_\T^0$, then the
\pts{} uncertainty in an $N$-jet event becomes $E_\T^0\sqrt N$. Since
$E_\T^0$ is 20 GeV (30 GeV) for $\mult=2000$ (3000), the \pts{} uncertainty
ranges from 28 GeV (42 GeV) for $N=2$ to 40 GeV (60 GeV) for $N=3$. This
background can easily be eliminated by observing distributions with a cut
$\pts>50$ GeV. As noted in the preceding paragraph, the number of events
is still very large.

A second important background comes from limited angular coverage of the
detector. Missing \pt{} could be generated by a jet lying outside the
detector. An ISAJET simulation showed that for a detector coverage of
$|y|\le5$ and all polar angles, the background to signal ratio can be
reduced to less than $10^{-5}$ by requiring that the observed jets all
lie within $|y|\le2$.

A third background comes from $W$ or $Z$ production with a subsequent
leptonic decay giving a neutrino which escapes undetected. This
background is only 4\% of the signal, since the cross section is damped
by the ratio of the weak and strong couplings times the semi-leptonic
branching ratio of the vector boson. It may be reduced even further by
vetoing on a hard lepton in the direction of \Pts. Kinematically
similiar backgrounds may arise from production of non-standard-model
particles. These are small, and one could explore ways of reducing
them through topological or kinematic cuts. All these backgrounds will
be studied in greater detail in a separate publication.

{\bf The suggested experiment:}\
For an LHC detector with full polar angle and $|y|\le5$ coverage, we suggest
the following measurement:
\item{1)} Select jet events with all jets harder than 50 GeV and $|y|\le2$,
  and bin according to the number of jets $N$.
\item{2)} Veto events with missing transverse momentum less than 50 GeV.
\item{3)} Veto events with one or more hard charged leptons in a
  cone $\Delta R=0.5$ around the direction of the missing momentum.
\item{4)} Match the $\pts$ distribution against the formul\ae{} in this
  letter, varying \ti{} for each $N$. The result is a measurement of \ti.

\noindent
About $10^5$ events are expected to pass the cuts in steps 1--3 for an
integrated luminosity of $10^4\;\mu{\rm b}^{-1}$. Typically 10\% of the
events will have $\pts>50$ GeV. If heavy-ion collisions give
rise to a quark-gluon plasma at the LHC, then in a year of
running one could expect to obtain a good measurement of the initial
temperature of the plasma.

I would like to thank Nirmalya Parua for help with ISAJET runs.

\vfill\eject
\bigskip\centerline{\sectnfont REFERENCES}\bigskip
\item{\reftag{appel})}
  D.~A.~Appel, \PR D 33 (1986) 717.
\item{\reftag{larry})}
  J.-P.~Blaizot and L.~D.~McLerran, \PR D 34 (1986) 2739.
\item{\reftag{heinz})}
  M.~Rammerstorfer and U.~Heinz, \PR D 41 (1990) 306.
\item{\reftag{gale})}
  J.~Pan and C.~Gale, McGill University preprint, McGill/94-16.
\item{\reftag{miklos})}
  M.~Gyulassy and M.~Pl\"umer, \PL B 243 (1990) 432;\lb
  M.~Gyulassy and M.~Pl\"umer, \NP B 346 (1990) 1;\lb
  M.~Gyulassy and X.-N.~Wang, \PRL 68 (1992) 1480;\lb
  M.~Gyulassy and X.-N.~Wang, \NP B 420 (1994) 583.
\item{\reftag{parisi})}
  G.~Parisi and R.~Petronzio, \NP B154 (1979) 427.
\item{\reftag{cs})}
  J.~C.~Collins and D.~E.~Soper, \NP B 197 (1982) 446;\lb
  J.~C.~Collins and D.~E.~Soper, \NP B 250 (1985) 199.
\vfil\end